\documentclass[sigconf,natbib=true,screen=true]{acmart}

\usepackage{acronym}
\usepackage[skip=0pt]{caption}
\usepackage[inline]{enumitem}
\usepackage{booktabs} 
\usepackage{multirow}
\usepackage{array}
\usepackage{pifont}
\usepackage{xspace}
\usepackage{subcaption}
\usepackage{makecell}
\usepackage{CJKutf8}
\usepackage{tcolorbox}
\usepackage[para]{footmisc}
\usepackage{amsfonts}
\usepackage{rotating}

\acrodef{QF-DC}{query filtering mechanism based on document relevance and conversation alignment}

\acrodef{RAG}{retrieval-augmented generation}

\acrodef{RCD}{Retrieval from Conversational Dialogues}
\acrodef{PSC}{proactive search in conversations}
\acrodef{CPS}{conversational proactive search}
\acrodef{PS}{proactive search}

\acrodef{GR}{generative retrieval}
\acrodef{LLM}{large language model}
\acrodef{QPP}{query performance prediction}
\acrodef{IR}{information retrieval}
\acrodef{NLP}{natural language processing}
\acrodef{PEFT}{parameter-efficient fine-tuning}
\acrodef{ICL}{in-context learning}
\acrodef{LoRA}{low-rank adaptation}

\acrodef{RR}{reciprocal rank}
\acrodef{AP}{Average Precision}
\acrodef{nDCG}{normalized discounted cumulative gain}

\acrodef{HSD}{Tukey's honestly significant difference}
\acrodef{ANOVA}{analysis of variance}

\acrodef{UEF}{utility estimation framework}
\acrodef{QPP-PRP}{pairwise rank preference-based QPP}
\acrodef{M-QPPF}{multi-task query performance prediction framework}
\acrodef{WRIG}{weighted relative information gain-based model}
\acrodef{NLP}{natural language processing}
\acrodef{CIS}{conversational information seeking}
\acrodef{ANCE}{Approximate nearest neighbor Negative Contrastive Estimation}

\acrodef{RL}{reinforcement learning}

\AtBeginDocument{%
  \providecommand\BibTeX{{%
    \normalfont B\kern-0.5em{\scshape i\kern-0.25em b}\kern-0.8em\TeX}}}

\newcommand{\our}{Conv2Query\xspace}

\makeatletter
\@ifpackagelater{acronym}{2020/01/01}
{%
}
{%

\newcommand{\Ac}[1]{\ac{#1}(\textbf{properly capitalized})}
\newcommand{\Acf}[1]{\acf{#1}(\textbf{properly capitalized})}
}%
\makeatother

\newcommand{\header}[1]{\vspace*{1mm}\noindent\textbf{#1}.}

\copyrightyear{2025}
\acmYear{2025}
\setcopyright{cc}
\setcctype{by}
\acmConference[SIGIR '25]{Proceedings of the 48th International ACM SIGIR Conference on Research and Development in Information Retrieval}{July 13--18, 2025}{Padua, Italy}
\acmBooktitle{Proceedings of the 48th International ACM SIGIR Conference on Research and Development in Information Retrieval (SIGIR '25), July 13--18, 2025, Padua, Italy}\acmDOI{10.1145/3726302.3729915}
\acmISBN{979-8-4007-1592-1/2025/07}

\makeatletter
\gdef\@copyrightpermission{
  \begin{minipage}{0.3\columnwidth}
   \href{https://creativecommons.org/licenses/by/4.0/}{\includegraphics[width=0.90\textwidth]{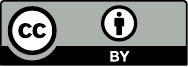}}
  \end{minipage}\hfill
  \begin{minipage}{0.7\columnwidth}
   \href{https://creativecommons.org/licenses/by/4.0/}{This work is licensed under a Creative Commons Attribution International 4.0 License.}
  \end{minipage}
  \vspace{5pt}
}
\makeatother

\author{Chuan Meng}
\orcid{0000-0002-1434-7596}
\affiliation{%
  \institution{University of Amsterdam}
  \city{Amsterdam}
  \country{The Netherlands}
}
\email{c.meng@uva.nl}

\author{Francesco Tonolini}
\orcid{0009-0002-1100-9556}
\affiliation{%
  \institution{Amazon}
  \city{London}
  \country{United Kingdom}
}
\email{tonolini@amazon.com}

\author{Fengran Mo}
\orcid{0000-0002-0838-6994}
\affiliation{%
  \institution{Université de Montréal}
  \city{Montréal}
  \country{Canada}
}
\email{fengran.mo@umontreal.ca}

\author{Nikolaos Aletras}
\orcid{0000-0003-4285-1965}
\affiliation{%
  \institution{Amazon \& University of Sheffield}
  \city{London}
  \country{United Kingdom}
}
\email{aletras@amazon.com}

\author{Emine Yilmaz}
\orcid{0000-0003-4734-4532}
\affiliation{%
  \institution{Amazon \& University College London}
  \city{London}
  \country{United Kingdom}
}
\email{eminey@amazon.com}

\author{Gabriella Kazai}
\orcid{0009-0002-5158-6630}
\affiliation{%
  \institution{Amazon}
  \city{London}
  \country{United Kingdom}
}
\email{gkazai@amazon.com}



\begin{document}

\title[Bridging the Gap: From Ad-hoc to Proactive Search in Conversations]{Bridging the Gap:\\ From Ad-hoc to Proactive Search in Conversations}

\renewcommand{\shortauthors}{Chuan Meng et al.}

\begin{abstract}
Proactive search in conversations (PSC) aims to reduce user effort in formulating explicit queries by proactively retrieving useful relevant information given conversational context. Previous work in PSC either directly uses this context as input to off-the-shelf ad-hoc retrievers or further fine-tunes them on PSC data. However, ad-hoc retrievers are pre-trained on short and concise queries, while the PSC input is longer and noisier. This input mismatch between ad-hoc search and PSC limits retrieval quality. While fine-tuning on PSC data helps, its benefits remain constrained by this input gap. In this work, we propose Conv2Query, a novel conversation-to-query framework that adapts ad-hoc retrievers to PSC by bridging the input gap between ad-hoc search and PSC. Conv2Query maps conversational context into ad-hoc queries, which can either be used as input for off-the-shelf ad-hoc retrievers or for further fine-tuning on PSC data. Extensive experiments on two PSC datasets show that Conv2Query significantly improves ad-hoc retrievers' performance, both when used directly and after fine-tuning on PSC.
\end{abstract}

\maketitle

\acresetall



\begin{figure}[ht]
    \centering
    \begin{subfigure}{\columnwidth}
    \includegraphics[width=\linewidth]{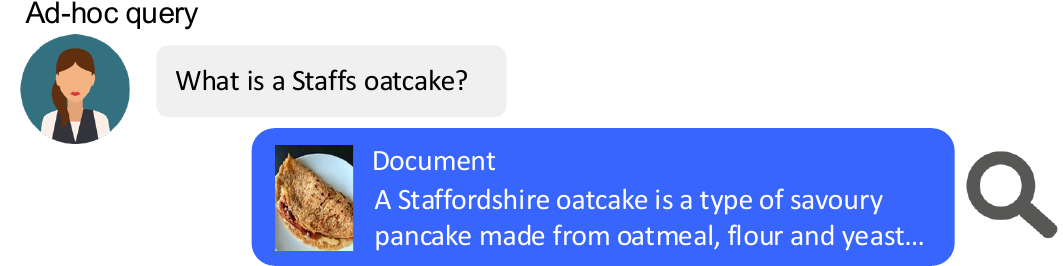}
        \caption{Ad-hoc search. Ad-hoc retrievers are typically trained on concise ad-hoc queries.
        }
        \label{fig:ad-hoc_definition}
    \end{subfigure}
    \begin{subfigure}{\columnwidth}
    \includegraphics[width=\linewidth]{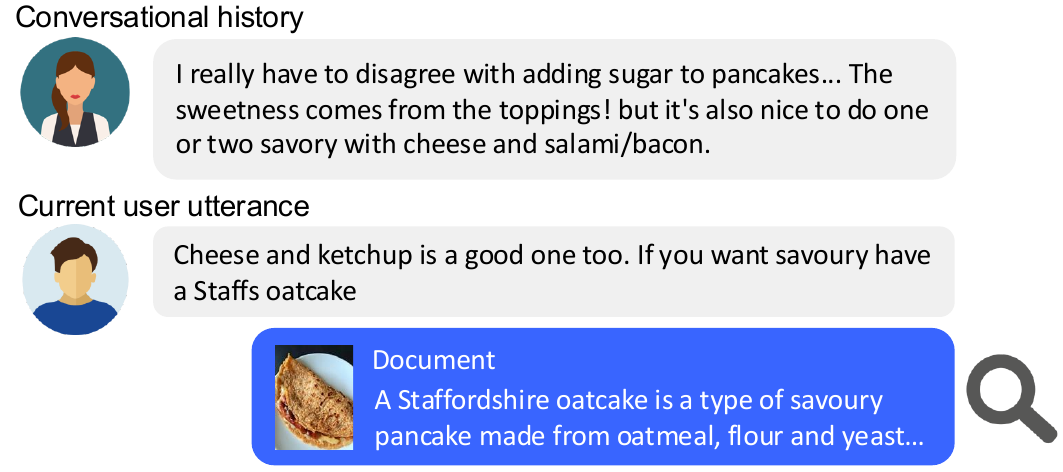}
        \caption{
        Proactive search in a multi-party conversation from the ProCIS dataset~\citep{samarinas2024procis}. 
        Two avatars represent two users in the conversation.
    Prior work typically inputs lengthy and noisy conversational context (e.g., concatenated history and the current utterance) into ad-hoc retrievers.
    } 
    \label{fig:psc_definition}
    \end{subfigure}
    %
    \caption{
    Comparison between ad-hoc search and proactive search in conversations.
    }
    \label{fig:definition}
    \vspace*{-1mm}
\end{figure}

\section{Introduction}
%
\Ac{PSC} aims to retrieve relevant documents based on an ongoing conversation without an explicit query from the user~\citep{samarinas2024procis,ros2023retrieving,pal2021effective,ganguly2020overview,andolina2018investigating}.
This contrasts with traditional ad-hoc or conversational search which typically follows the popular ``query–res\-ponse''~\citep{lai2024adacqr,yoon2024ask,ye2023enhancing,jang2023itercqr} or ``query–clarifi\-cation'' paradigms~\citep{lu2025zero,zhang2024ask,butala2024promise,meng2023system}, where users issue explicit queries, and then the system retrieves information or asks clarifying questions.
%
\ac{PSC} has been shown to not only reduce user effort in formulating and refining explicit queries in conversations, but also enrich conversations by proactively introducing relevant facts and ideas~\citep{andolina2018investigating}.
%
Specifically, \ac{PSC} supports conversations in two key ways:
\begin{enumerate*}[label=(\roman*)]
\item \textit{Conversation contextualisation}~\citep{samarinas2024procis,pal2021effective,ganguly2020overview}. 
\ac{PSC} can proactively retrieve relevant documents to clarify concepts or fact-check claims, before the user explicitly asks for them~\citep{samarinas2024procis,pal2021effective,ganguly2020overview}.
E.g., Figure~\ref{fig:psc_definition}, shows a conversation about ``pancake'', where a user mentions a specific pancake ``Staffs oatcake.'' 
The search system returns a document explaining ``Staffordshire oatcake,'' helping the user understand it without needing to search manually (e.g., ``What is a Staffs oatcake?'' in Figure~\ref{fig:ad-hoc_definition}).
\item \textit{Interest anticipation}~\citep{samarinas2024procis,ros2023retrieving}. 
\ac{PSC} can proactively retrieve documents aligning with users' next potential interests before they ask for them, also known as \textit{interest anticipation}~\citep{samarinas2024procis,ros2023retrieving}.
As shown in Figure~\ref{fig:psc_definition}, even without the current user utterance, after a user states, ``it's also nice to do one or two savoury,'' the system proactively retrieves information about ``Staffordshire oatcake,'' a savoury pancake. 
By retrieving this information in advance, the system eliminates the need for a manual query (e.g., ``What pancakes are savoury?'') and ensures a natural conversation flow.
\end{enumerate*}

\header{Motivation}
%
Existing studies typically feed raw conversational context into off-the-shelf ad-hoc lexical/neural retrievers, or further fine-tune the neural ones on \ac{PSC} using this raw context~\citep{samarinas2024procis,ros2023retrieving}.
However, previous work faces three key limitations:
\begin{enumerate*}[label=(\roman*)]
\item \textit{Input gap between ad-hoc pre-training and \ac{PSC} inference}.
Directly feeding raw conversational context into neural retrievers pre-trained on ad-hoc search data (e.g., MS MARCO~\citep{bajaj2016ms}) leads to poor retrieval quality~\citep{ros2023retrieving}. 
This is due to the mismatch between the input format used in pre-training and inference: ad-hoc neural retrievers are pre-trained on short and concise queries, whereas in \ac{PSC}, they receive longer and noisier conversational contexts.
Such a distribution shift between training and inference hinders retrieval effectiveness~\citep{zhuang2022bridging,rau2022role,zhuang2021dealing}.
\item \textit{Input gap between ad-hoc pre-training and \ac{PSC} fine-tuning}. 
While further fine-tuning ad-hoc neural retrievers on PSC data improves performance~\citep{samarinas2024procis,ros2023retrieving}, the retrieval quality might still be limited by the input mismatches between the source ad-hoc search task (ad-hoc queries) and the target \ac{PSC} task (conversational context).
The discrepancy limits neural retrievers' ability to fully leverage pre-trained ad-hoc knowledge, hindering effective transfer learning~\citep{shi2024continual,zhang2022survey}.
\item \textit{Limited performance of lexical retrievers}.
Prior work has shown that traditional lexical retrievers (e.g., BM25 \citep{robertson1995okapi}) struggle with verbose conversational contexts~\citep{samarinas2024procis,ros2023retrieving}.
Unlike neural retrievers, they cannot be fine-tuned on \ac{PSC} data.
\end{enumerate*}


\header{A novel framework for \ac{PSC}}
%
%
To tackle the above limitations, we propose a \textit{Conversation-to-Query framework} (\our) for \ac{PSC}, which aims to effectively adapt ad-hoc neural retrievers to \ac{PSC}.
\our aims to transform lengthy, noisy conversational context into short, concise ad-hoc queries that closely resemble the format of queries in widely-used ad-hoc search datasets, like MS MARCO~\citep{bajaj2016ms}); these datasets have been widely-used for training numerous state-of-the-art retrievers~\citep{ma2024fine,formal2022distillation,formal2021splade,xiong2021approximate}.
We hypothesise that \our can effectively improve the performance of ad-hoc neural retrievers on \ac{PSC} by providing ad-hoc queries during both inference and fine-tuning on \ac{PSC} data; and \our can also effectively improve lexical retrievers' performance by delivering concise queries that eliminate noises in conversational contexts.

\header{Learning pseudo ad-hoc query targets}
Modelling \our poses a key challenge.
In \ac{PSC}, users' search intents for each conversational context are implicit, and we need to generate ad-hoc queries that reveal the implicit search intent for each context.
However, no ground-truth ad-hoc query training targets revealing implicit search intents are available to optimise the mapping in \our. 
Also, our preliminary experiments show that directly prompting \acp{LLM} to generate ad-hoc queries from verbose conversational context yields limited retrieval performance.
To overcome the issue, we propose to generate pseudo ad-hoc query training targets for \our from annotated relevant documents for each conversational context.
Because users' implicit search intents are well reflected in the relevant documents, the query targets generated from relevant documents have the potential to capture the implicit search intents.
To perform the document-to-query mapping process, we leverage a Doc2Query model~\citep{gospodinov2023doc2query,nogueira2019doc2query} because it has been pre-trained on ad-hoc search data to take a document as input and generate ad-hoc queries that the document might answer.
Following prior work~\citep{gospodinov2023doc2query}, we first use Doc2Query to generate a set of queries for each relevant document, and then use query--document relevance filtering to select the optimal query target.
The filtering ensures that the selected target is highly relevant to the relevant document for a conversational context.

\header{A new query filtering mechanism for \ac{PSC}}
We find that query--document relevance filtering often creates a semantic gap between pseudo ad-hoc query targets and their conversational context, making it harder for the \ac{LLM} to learn an effective mapping. 
E.g., in the case illustrated in Figure~\ref{fig:psc_definition}, a query like ``What is a Staffordshire oatcake?'' ranks high in query–document relevance but lacks semantic alignment with the conversational context.
In contrast, a query like ``What are recipes for savoury oatcakes?'' is both relevant to the document and aligned with the conversation.
However, query–document relevance filtering tends to overlook such contextually aligned queries.
To address this issue, we propose \textit{QF-DC}, a query filtering mechanism that selects optimal query targets based on both document relevance and contextual alignment.

%



\header{Experiments}
We find that \our significantly improves the performance of reusing off-the-shelf ad-hoc lexical/neural retrievers on \ac{PSC}~(Section~\ref{sec:gap1}).
\our enables off-the-shelf ad-hoc neural retrievers to achieve retrieval quality on par with or better than the ones fine-tuned on \ac{PSC} with raw context.
Also, \our improves the performance of ad-hoc neural retrievers after further fine-tuning them on \ac{PSC}~(Section~\ref{sec:gap2}).
Furthermore, we assess the impact of query filtering (Section~\ref{sec:filter}), indicating that QF-DC leads to faster convergence and higher retrieval quality.
Moreover, we find that \our performs consistently well in various \ac{LLM} configurations~(Section~\ref{sec:llm}).


\header{Contributions}
Our main contributions are as follows:
\begin{itemize}[leftmargin=*,nosep]
\item We propose \our for \ac{PSC}, which effectively adapts ad-hoc retrievers to \ac{PSC} by bridging the input gap between ad-hoc pre-training and \ac{PSC} fine-tuning/inference.
\item We devise a query filtering mechanism (QF-DC) that selects optimal pseudo ad-hoc query targets based on both document relevance and alignment with conversational context.
\item Experimental results show that \our significantly improves the performance of reusing off-the-shelf ad-hoc retrievers and their performance after fine-tuning on \ac{PSC}. 
We release our code and data at~\url{https://github.com/ChuanMeng/Conv2Query}.
\end{itemize}

\section{Related work}

\subsection{Proactive search in conversations}
\subsubsection{Proactive search}
Unlike traditional search following a ``query--response'' paradigm, where users issue explicit queries and the system retrieves information, \ac{PS} is query-free, which aims to retrieve relevant information without the user explicitly submitting a query~\cite{jones2022workshop}. 
%
%
Because there is no explicit user query, \ac{PS} utilises users' latent information needs inferred from users' contexts to perform search.
\Ac{PS} has been explored by using various users' contexts, including but not limited to
users' historical queries (past queries issued by the user, or similar queries issued by other users)~\citep{sen2021know,sen2018procrastination,elliott2009proactive},
web pages browsed by users~\citep{liebling2012anticipatory,cheng2010actively},
information about tasks being performed by users~(e.g., documents/emails they are reading or writing)~\citep{koskela2018proactive,vuong2017watching,vuong2017proactive,tran2016forgotten,dumais2004implicit,rhodes2000just,rhodes1996remembrance},
physical attributes (e.g., time or location)~\citep{song2016query,yang2016modeling,shokouhi2015queries,fitchett2012accessrank},
query description~\citep{ahmed2022towards}, 
ongoing stream of TV broadcast news~\citep{henzinger2003query}, 
user status text in social media~\citep{park2016mobile},
and
conversational context~\citep{samarinas2024procis,ros2023retrieving}.

\subsubsection{Proactive search in conversations}
\Acf{PSC} has recently gained increased attention but remains under-explored~\citep{samarinas2024procis,ros2023retrieving,ganguly2020overview}.
\citet{andolina2018investigating} carry out a user study to show the benefits of \ac{PSC}.
In the study, participants have conversations while accessing a proactive search system that monitors conversation, detects entities mentioned in the conversation and proactively retrieves and presents documents/entities relevant to the conversation.
They found that \ac{PSC} supports conversations with facts/ideas, and reduces users' effort needed to formulate and refine explicit queries.
A key bottleneck in developing \ac{PSC} models is the lack of benchmarks.
\citet{ganguly2020overview} introduce the \ac{RCD} benchmark for \ac{PSC}, which encourages research on developing systems capable of proactively retrieving relevant documents to contextualise hard-to-understand concepts within a conversation.
On this dataset, \citet{pal2021effective} propose to identify text windows that are likely to be hard-to-understand concepts in conversations, and then perform retrieval based on the identified text windows.
However, \ac{RCD} is limited by its small scale and reliance on movie scripts for conversations, making it less realistic.
Furthermore, \citet{ros2023retrieving} introduce the WebDisc dataset, a larger and more realistic alternative.
It uses Reddit threads as conversations, where users include hyperlinks to external webpages.
These hyperlinks often serve as citations or provide additional context, supporting ongoing discussions. 
Recognising this, \citet{ros2023retrieving} regard the linked webpages as relevant documents tied to Reddit threads.
Recently, \citet{samarinas2024procis} curate the ProCIS dataset using a similar approach. 
ProCIS stands out with a larger corpus and training dataset. 
%
On both datasets, \citet{samarinas2024procis,ros2023retrieving} feed the raw conversational context into either a lexical retriever or a neural retriever pre-trained on ad-hoc search data.
They also further fine-tune these neural retrievers on \ac{PSC} training data before retrieval.

Our work differs from these studies, as we propose to transform conversational context into ad-hoc queries before using ad-hoc lexical/neural retrievers.

\subsection{Query prediction}

Our work is related to two key research directions in query prediction: Doc2Query and next query prediction.
\subsubsection{Doc2Query}
Given a source document, Doc2Query is a process of generating queries that the document might answer~\citep{nogueira2019document}.
Doc2Query has demonstrated benefits in document expansion~\citep{basnet2024deeperimpact,gospodinov2023doc2query,nogueira2019doc2query} and synthetic data generation~\citep{jeronymo2023inpars,dai2023promptagator,bonifacio2022inpars}.
In the former, Doc2Query generates a set of relevant queries for each document and appends them to the document before indexing; this expansion improves retrieval quality by bridging term mismatches between user queries and relevant documents.
In the latter, Doc2Query generates queries likely relevant to a given document, where each query--document pair forms a positive training example used to train traditional neural ranking models~\citep{chandradevan2024duqgen,jeronymo2023inpars,dai2023promptagator,bonifacio2022inpars} or generative retrieval models~\citep{zeng2024scalable,zhuang2022bridging}.
Additionally, Previous work~\citep{gospodinov2023doc2query,jeronymo2023inpars} found that some generated queries are irrelevant to the source document, potentially hurting the performance of downstream applications using these queries.
To address this issue, these studies~\citep{gospodinov2023doc2query,jeronymo2023inpars} use a query filtering mechanism to remove irrelevant generated queries based on their relevance to the source document.

Our work differs from previous studies in three key aspects:
\begin{enumerate*}[label=(\roman*)]
\item we propose \our that generates ad-hoc queries from conversational context instead of documents;
\item we leverage Doc2Query to create training data of ad-hoc queries, enabling \our to learn the mapping from conversation context to ad-hoc queries; and
\item we propose a query filtering mechanism based on \textit{query--conversation relevance}, ensuring that our ad-hoc query pseudo labels align closely with both target document and conversational context.
\end{enumerate*}


\subsubsection{Query suggestion}
Query suggestion (a.k.a query recommendation) is a core task in session search~\citep{li2012dqr,bonchi2012efficient,boldi2008query}.
It aims to predict the next user query given past users’ search behaviours.
Query suggestion can assist users in formulating their queries~\citep{rosset2020leading}, which is particularly valuable when an information need requires multiple searches~\citep{feild2013task}.
Specifically, query suggestion has been studied to predict the next query based on various types of information, such as user historical queries in the current or previous search sessions~\citep{rosset2020leading,chen2018attention,dehghani2017learning,sordoni2015hierarchical,muntean2013learning}, user feedback on the search result (such as browsing and clicks)~\citep{wu2018query}, or pre-search context (e.g., the news article a user browsed before the search)~\citep{kong2015predicting}.

Unlike query suggestion, which predicts the next query based on session search data (e.g., query logs or clickthrough data), our work is to generate queries directly from noisy and verbose conversational context, without an explicit user query at the moment.
Additionally, it is important to note that \citet{yang2017neural} focus on selecting the next question a user might ask in a conversation from a predefined pool of user question candidates. 
We differ as we generate queries from the conversational context without relying on an existing set of user question candidates.


%



\subsection{Conversational search}
Conversational search aims to retrieve relevant documents for users' context-dependent queries in a multi-turn conversation~\citep{mo2025conversational,abbasiantaeb2025improving,mo2024survey}.
These context-dependent queries often contain omissions, coreferences or ambiguities, making it difficult for ad-hoc search methods to capture the underlying information need.
Two main research directions address the context-dependent query understanding problem: conversational query rewriting and conversational dense retrieval.
Conversational query rewriting aims to transform context-dependent queries into self-contained ones~\citep{mo2023convgqr,mo2024chiq,lai2024adacqr,yoon2024ask,ye2023enhancing,mao2023large,jang2023itercqr,yu2020few,mo2023learning,mao2023search}, while conversational dense retrieval trains a query encoder to encode the current user query and conversational history into a contextualized query embedding that is expected to implicitly represent the information need of the current query in a latent space~\citep{mao2024chatretriever,hai2023cosplade,lin2021contextualized,yu2021few,mo2024history,mo2024aligning}.

The key difference between \ac{PSC} and conversational search is that conversational search has explicit user query at each turn, whereas \ac{PSC} operates on conversation context alone, without a current explicit user query.
%
Note that we do not use conversational query rewriting methods as baselines in this paper, because there is no explicit user query in \ac{PSC} for rewriting, making these methods inapplicable to \ac{PSC}.

%
%


\subsection{Proactive response prediction}
Proactive conversational response prediction aims to produce a system response that guides the conversation direction~\citep{deng2024towards,deng2023survey,liao2023proactive1,liao2023proactive2,deng2023rethinking}.
Various types of proactive response prediction have been explored, such as clarifying question prediction~\citep{zhang2024ask,butala2024promise,deng2022pacific,deng2023prompting}, user preference elicitation~\citep{zhang2018towards}, persuasion~\citep{mishra2022pepds}, target-steering~\citep{wang2023target,zhu2021proactive}, item recommendation~\citep{sun2018conversational}, suggesting follow-up questions~\citep{butala2024promise,tayal2024dynamic,lee2024redefining,yan2018smarter}, and providing additional information~\citep{lee2024redefining,balaraman2020proactive}.
Amongst these, providing additional information is most relevant to \ac{PSC}, which aims to proactively produce a response offering supplementary and useful information not explicitly requested by users~\citep{lee2024redefining,balaraman2020proactive}.
For example, a recent study~\citep{lee2024redefining} prompt \acp{LLM} to generate a proactive response that consists of the answer to the user’s query and a proactive element, which refers to new information related to the initial query.
However, instead of focusing on response generation, \ac{PSC} focuses on retrieving relevant documents to offer additional information to users in the absence of an explicit user query.






\section{Task definition}

Given a conversational context $C_t$ at turn $t$ and a corpus of documents $D = \{d_1, d_2, \cdots, d_{|D|}\}$, the goal of \ac{PSC} is to develop a ranking model that retrieves a ranked list of $k$ documents $D_t = \{d_{t,1}, d_{t,2}, \cdots, d_{t,k}\}$ from $D$; $D_t$ provides relevant information (e.g., facts or ideas) to support $C_t$~\citep{samarinas2024procis}.
Note that a user utterance in the conversational context $C_t$ can take any form and is not necessarily a query.
%
%
Following~\citet{ros2023retrieving}, we study two settings:
\begin{enumerate*}[label=(\roman*)]
\item \textit{Conversation contextualisation} (referred to as the ``full'' setting in \citep{ros2023retrieving}): $C_t$ consists of user utterances from turns $1$ to $t$, including the conversational history $\{u_1, u_2, \dots, u_{t-1}\}$ (user utterances up to turn $t-1$) and the current user utterance $u_t$ at turn $t$.
The goal is to retrieve relevant documents $D_t$ to clarify hard-to-understand concepts mentioned in $u_t$ or to verify factual claims made by the user in $u_t$.
\item \textit{Interest anticipation} (referred to as the ``proactive'' setting in \citep{ros2023retrieving}): $C_t$ consists of conversational history $\{u_1, u_2, \cdots, u_{t-1}\}$ with user utterances up to turn $t-1$. 
The aim is to retrieve documents $D_t$ aligned with the user’s interest at turn $t$.
In other words, the ranking model $f$ must anticipate the information the user is likely to explore at turn $t$, based on the conversational history up to turn $t-1$.
\end{enumerate*}
\citet{ros2023retrieving} has shown that this setting is more challenging than conversation contextualisation.

\citet{ros2023retrieving} also explore the ``last'' setting, where $C_t$ includes only the current user utterance $u_t$; they found that retrievers perform similarly whether they use only the current user utterance $u_t$ or combine it with the conversational history.
We exclude this setting, as we believe only using the current utterance is insufficient for practical applications that often require context from prior interactions.
Additionally, \citet{samarinas2024procis}consider a ``reactive'' setting,  where retrieval occurs only after a conversation reaches its final turn $T$.
We do not adopt this setting, as it not suitable for delivering timely information to support ongoing conversations.

\section{Method}
\label{sec:model}

\our consists of five phases:
\begin{enumerate*}[label=(\roman*)]
\item generating ad-hoc queries from documents,
\item query filtering by relevance to documents and conversations,
\item learning to generate ad-hoc queries from conversations,
\item at inference, generating ad-hoc queries for retrieval, and
\item (optionally) fine-tuning ad-hoc retrievers via filtered ad-hoc queries.
\end{enumerate*}
Note that phases (i) and (ii) correspond to training data preparation for \our, (iii) concerns the training of \our, and (iv) pertains to inference.

In (i), we leverage a Doc2Query model to generate $n$ ad-hoc query candidates from a document (see Section \ref{sec:method_doc2query}); 
in (ii), we introduce a novel query filtering mechanism (QF-DC) that evaluates query--document relevance and query--conversation alignment to select the optimal ad-hoc query target that is relevant to both its source documents and conversational context (see Section \ref{sec:method_qf}).
In (iii), we fine-tune \our model to learn the mapping from a conversational context to its filtered ad-hoc query (see Section \ref{sec:learning}).
(iv), at inference, given a conversational context, we generate an ad-hoc query to be used with any ad-hoc retriever (see Section \ref{sec:infer}).
(v) is optional: we fine-tune an ad-hoc neural retriever on \ac{PSC} by using our  filtered ad-hoc queries produced in (ii) (see Section \ref{sec:further_ft}).

\subsection{Generating ad-hoc queries from documents}
\label{sec:method_doc2query}

For a conversational turn $t$ annotated with a relevant document $d_t^{+}$, we leverage a Doc2Query~\citep{basnet2024deeperimpact,nogueira2019doc2query} model to map $d_t^{+}$ to a set of $n$ ad-hoc query candidates that $d_t^{+}$ might answer.
Formally,
\begin{equation}
\begin{split}
 \{q_{t,1}, q_{t,2}, \dots, q_{t,n}\} = \mathrm{f_{Doc2Query}}(d_t^{+}),
\end{split}
\label{eq:doc2query}
\end{equation}
where $\{q_{t,1}, q_{t,2}, \dots, q_{t,n}\}$ represent $n$ generated ad-hoc query candidates.
Doc2Query models excel at generating ad-hoc queries from documents due to its pre-training on query--document pairs from MS MARCO~\citep{bajaj2016ms}, a widely-used ad-hoc search dataset; the model architecture can be based on a language model (e.g., T5~\citep{gospodinov2023doc2query,nogueira2019doc2query}, Llama 2~\citep{basnet2024deeperimpact}).

\subsection{Query filtering by relevance to documents and conversations}
\label{sec:method_qf}

We propose a query filtering mechanism that evaluates both query--document relevance and query--conversation relevance, ensuring the selection of queries that are highly relevant to their source document while also align with the corresponding conversational context. 
Specifically, given the generated $n$ ad-hoc query candidates $\{q_{t,1}, q_{t,2}, \dots, q_{t,n}\}$, we select an optimal query candidate $q_t^*$ by identifying the query candidate with the highest aggregated score across all $n$ candidates. 
This score is derived by aggregating the query--document relevance and query--conversation relevance scores.
%
Formally,
\raggedbottom
\begin{equation}
\begin{split}
i^* =& \arg\max_{i \in \{1, \dots, n\}} s_{t,i}, \\
q_t^* =& q_{t,i^*}, \\
s_{t,i} =& \mathrm{f_{aggregate}}(s^{qd}_{t,i},s^{qc}_{t,i})\in\mathbb{R},
\end{split}
\label{eq:filter_overall}
\end{equation}
where \( s_{t,i} \) represents the aggregated score for the $i$-th query candidate $q_{t,i}$, calculated using the aggregation function $\mathrm{f_{aggregate}}(\cdot, \cdot)$ (e.g., summation). 
\smash{$s^{qd}_{t,i}$} and \smash{$s^{qc}_{t,i}$} denote the query--document relevance score and the query--conversation relevance score for $q_{t,i}$, respectively. 
We follow~\citet{gospodinov2023doc2query} to compute the query--document relevance score \smash{$s^{qd}_{t,i}$}:
\begin{equation}
\begin{split}
s^{qd}_{t,i} = \mathrm{f_{relevance}}(q_{t,i},d_t^{+}) \in\mathbb{R},
\end{split}
\label{eq:filter_qd}
\end{equation}
where $\mathrm{f_{relevance}}(\cdot,\cdot)$ is a relevance prediction model that maps a query--document pair to a relevance score, with higher scores representing greater relevance.
$\mathrm{f_{relevance}}(\cdot,\cdot)$ can be either a re-ranker (e.g., MonoT5~\citep{nogueira2020document}) or a retriever~(e.g., TCT-ColBERT~\citep{lin2021batch}).
We calculate the query--conversation relevance score \smash{$s^{qc}_{t,i}$} in a similar way:
\begin{equation}
\begin{split}
s^{qc}_{t,i} = \mathrm{f_{relevance}}(q_{t,i},C_t) \in\mathbb{R},
\end{split}
\label{eq:filter_qc}
\end{equation}
where the value of $s^{qc}_{t,i}$ is higher if $q_{t,i}$ is more relevant to its corresponding conversational context.

\begin{figure}[t]
     \centering
     \begin{tcolorbox}[notitle,boxrule=1pt,colback=gray!5,colframe=black, arc=2mm,width=\columnwidth]
\sf\textbf{Instruction}: Based on the following conversation history and the current user utterance, please generate a search query that retrieves documents relevant to the current user utterance.\\
    Conversational history: \{\} \\
    Current user utterance: \{\} \\
    Generated query:
    \end{tcolorbox}
     \caption{
    Prompt for conversation contextualisation.
     }
     \label{fig:prompt-context}
\end{figure}

\subsection{Learning to generate ad-hoc queries from conversations}
\label{sec:learning}

We treat the selected ad-hoc query $q_t^{*}$ as a learning target and pair it with the corresponding conversational context $C_t$ to form a training data point ($C_t$,$q_t^{*}$).
It enables us to train our \our model in mapping from a conversational context $C_t$ to $q_t^{*}$, represented as $C_t\rightarrow q^*_t$.
The loss function for a conversation with $T$ turns is defined as follows:
\begin{equation}
\label{eq:loss}
\begin{split}
\mathcal{L}(\theta_{\our}) = -\frac{1}{Z} \sum_{t \in \{t \mid I(t) = 1\}}^{T} \log P(q_t^{*} \mid \mathrm{prompt}(C_t)),
\end{split}
\end{equation}
where 
$I(t)$ is an indicator function that equals to 1 if turn $t$ is annotated with a relevant document and 0 otherwise. 
$Z = \sum_{i=1}^{T}I(t)$.
$\mathrm{prompt}(\cdot)$ is a prompt to instruct the \our model.

\subsection{Generating ad-hoc queries for retrieval}
\label{sec:infer}

At inference time, the trained \our model takes a conversational context $C_t $ at turn $t$ as input and generates an ad-hoc query $q'_t$.
This query $q'_t$ is then passed to a retrieval system, which returns a ranked list of relevant documents $D_t$:
\begin{equation}
\begin{split}
 q'_t =&\mathrm{f_{Conv2Query}}(\mathrm{prompt}(C_t)), \\
 D_t=&\mathrm{f_{retriever}}(q'_t).
\end{split}
\label{eq:infer}
\end{equation}

\subsection{Retriever fine-tuning using pseudo queries}
\label{sec:further_ft}

For each conversational context $C_t$, we pair its selected optimal ad-hoc query $q_t^*$ (Section \ref{sec:method_qf}) with the corresponding relevant document $d_t^{+}$ to create a positive training example ($q_t^*$, $d_t^{+}$).
We then sample negative documents for $C_t$ following standard neural retrieval practices and use both positive and negative examples to fine-tune a specific ad-hoc retriever on \ac{PSC}.

\begin{figure}[t]
     \centering
     \begin{tcolorbox}[notitle,boxrule=1pt,colback=gray!5,colframe=black, arc=2mm,width=\columnwidth]
\sf\textbf{Instruction}: Based on the following conversation history, please generate a search query that retrieves documents relevant to the next expected utterance.\\
    Conversational history: \{\} \\
    Generated query:
    \end{tcolorbox}
     \caption{
    Prompt for interest anticipation.
     }
     \label{fig:prompt-antici}
\end{figure}

\vspace*{-2mm}
\section{Experimental setup}
\label{sec:setup}

\header{Research questions}
Our work is steered by the following research questions:
\begin{enumerate}[label=\textbf{RQ\arabic*},leftmargin=*]
\item To what extent does \our bridge the input gap between ad-hoc pre-training and \ac{PSC} inference under conversation contextualisation and interest anticipation settings? \label{RQ1} \looseness=-1
\item How well \our bridge the input gap between ad-hoc pre-training and \ac{PSC} fine-tuning under the two settings? \label{RQ2} 
\item To what extent does our proposed query filtering mechanism (QF-DC) improve \our's performance? \label{RQ3}
\item To what extent the choice of \acp{LLM} impact \our's performance? \label{RQ4}
\end{enumerate}


\header{Datasets}
We use two recent large-scale datasets for proactive search in multi-party conversations: ProCIS \citep{samarinas2024procis} and WebDisc \citep{ros2023retrieving}:
\begin{itemize}[leftmargin=*,nosep]
\item {\bf ProCIS}~\citep{samarinas2024procis} consists of Reddit threads where multiple users engage in discussions; each conversation (thread) includes at least one utterance (comment) that contains Wikipedia hyperlinks; the hyperlink is added by the user when posting the comment.
These user-added Wikipedia articles serve as retrieval targets (sparse relevance judgments) because they frequently offer additional context or background information relevant to the ongoing conversation.
The dataset has a corpus of 5,315,384 Wikipedia articles; the average article length is 145.88 tokens (Llama tokenizer).
The dataset has four subsets: train, dev, future-dev, and test, containing 2,830,107, 4,165, 3,385, and 100 conversations (threads), respectively. The average number of turns per conversation in these subsets is 5.41, 4.91, 4.48, and 4.49.
The future-dev set only contains conversations that occur chronologically after those in the training set; so it can be used to evaluate a retrieval model's ability to generalise to newly emerging concepts and topics not seen during training.
The test set contains turn-level human-annotated dense relevance judgments, while other sets only has turn-level sparse relevance judgments based on user-included Wikipedia articles.
On the test set, each turn with associated relevant documents has an average of 2.30 relevant documents.

\item {\bf WebDisc}~\citep{ros2023retrieving} is built in a similar way to ProCIS, consisting of Reddit threads with turn-level sparse relevance judgments derived from user-added webpage hyperlinks.
This dataset has a corpus of 98,231 webpages not limited to Wikipedia.
\citet{ros2023retrieving} truncate overly long webpages to ensure compatibility with passage ranking models.
The dataset is split into train, validation and test sets, containing 128,404, 15,344 and 15,249 turns with user-added webpages, respectively.
\end{itemize}

\noindent For both datasets, raw links are already removed from Reddit utterances.
In a conversation (i.e., thread), only certain user utterances (i.e., comments) are associated with hyperlinks or human-labelled documents. 
Thus, the assumption is that a \ac{PSC} retriever should perform retrieval at those turns.





\header{Baselines}
We use retrievers under two settings: off-the-shelf ad-hoc retrievers or these further fine-tuned on \ac{PSC}.

For off-the-shelf ad-hoc retrievers, we evaluate them by feeding them three types of input without any fine-tuning on \ac{PSC} data.
%
First, following \citet{ros2023retrieving}, we feed conversational context into off-the-shelf ad-hoc retrievers.
Specifically, we use one lexical retriever BM25~\citep{robertson1995okapi}, and three neural retrievers that are pre-trained on the widely-used ad-hoc search dataset MS MARCO~\citep{bajaj2016ms}.
For the neural ones, we consider one learned sparse retriever, SPLADE++~\citep{formal2022distillation} (\texttt{splade-cocondenser-ensembledistil}), and two dense retrievers: ANCE~\citep{xiong2021approximate} (\texttt{ance-msmarco-passage}), and RepLLaMA~\citep{ma2024fine} (\texttt{repllama-v1-7b-lora-passage}), an \ac{LLM}-based state-of-the-art retriever.
Second, we consider two methods specifically designed for \ac{PSC}, which first process the conversational context before using off-the-shelf ad-hoc retrievers:
\begin{enumerate*}[label=(\roman*)]
\item Text Window~\citep{pal2021effective} first extracts text segments with size $k$ for the conversational context, and selects those likely to achieve high retrieval quality.
Following~\citet{pal2021effective}, we use the \ac{QPP} method NQC~\citep{shtok2012predicting} to estimate retrieval quality and set \(k=5\). However, we replace the original LM-Dirichlet~\citep{zhai2001study} retriever with the more recent and effective RepLLaMA.
\item LMGR~\citep{samarinas2024procis} first uses an \ac{LLM} to generate $n$ text descriptions for conversational context; then it retrieves $k$ documents per description, and use an \ac{LLM} to select the document that best matches each description.
We follow~\citet{samarinas2024procis} in using OpenChat-3.5 (enhanced Mistral-7B) as the \ac{LLM}, with \(n=20\) and \(k=5\), but replace the original ANCE retriever with the more recent RepLLaMA for consistency.
\end{enumerate*}
Third, we assess a \our variant that relies solely on prompting (only using Equation~\ref{sec:infer}). 
It directly prompts an \ac{LLM} to generate ad-hoc queries from conversational context.
We apply 1-shot and 2-shot prompting, denoted as \our-1-S and \our-2-S. 
\footnote{We randomly sample one or two training examples respectively, each consisting of a conversational context and its ad-hoc query selected by QF-DC from Doc2Query candidates.}
The queries generated by \our-1-S/-2-S are then fed into RepLLaMA for retrieval.

Regarding the ad-hoc retrievers further fine-tuned on \ac{PSC}, we use the three neural retrievers, ANCE~\citep{xiong2021approximate}, SPLADE++~\citep{formal2022distillation} and RepLLaMA~\citep{ma2024fine}, each fine-tuned on \ac{PSC} data.
All are fed with conversational context during fine-tuning on \ac{PSC}. 
%


\header{Evaluation metrics}
We follow \citet{ros2023retrieving} to use Precision@1 (P@1) and MRR@10 as our evaluation metrics for dev/val sets of both datasets.
This choice is made because these sets only contain sparse relevance judgments. 
Moreover, as \ac{PSC} is designed for conversational scenarios, it prioritises retrieving the most relevant document at the top of the ranked list. 
Therefore, precision-oriented metrics like Precision and MRR are particularly suitable.

For the ProCIS test set, which contains dense relevance judgments, we further use npDCG, a metric proposed by \citet{samarinas2024procis} specifically tailored to \ac{PSC}.
Unlike nDCG~\citep{jarvelin2002cumulated}, npDCG has three features:
\begin{enumerate*}[label=(\roman*)]
\item it aggregates DCG/iDCG values across all turns per conversation into a single score;
\item it avoids rewarding a retrieval model for returning the same relevant document across multiple turns; and
\item it evaluates the timing prediction of a \ac{PSC} system: a retriever can gain only when retrieving at turns with annotated judgments; the retriever incurs no gain/cost if it skips retrieval at a turn or retrieves at a turn without any annotated judgments.
\end{enumerate*}
Because each turn with associated judgements has 2.30 relevant documents on average, we use a cut-off of 5 for npDCG.
%
Note that our work focuses only on what to retrieve, and leaves retrieval timing prediction for future work.
So we assume perfect timing prediction and perform retrieval only at turns with annotated documents.

\header{Implementation details}
We perform BM25 retrieval via \texttt{Pyserini}; for ProCIS, we set \(k_1 = 0.9\) and \(b = 0.4\); for WebDisc, we follow \citet{ros2023retrieving} to set \(k_1 = 8, b = 0.99\) for the conversation contextualisation setting, and \(k_1 = 7, b = 0.99\) for the interest anticipation setting.
For all neural retrievers fed with conversational context, we observe that the default truncation length (e.g., 32 tokens) of the query encoders used on MS MARCO substantially reduces retrieval quality, as lots of important information in the lengthy conversational context is truncated. 
Thus, we set the truncation length for query encoders to 512, which exceeds the average conversational context length.
To further fine-tune a neural retriever on \ac{PSC}, we randomly sample negative documents from a mix of top 200 hard negatives retrieved by BM25 using the conversational history alone and the history combined with the user's current utterance, increasing negative diversity.
All neural retrievers use the same negative samples during fine-tuning on \ac{PSC} data.

%
Regarding our method, for Doc2Query \footnote{We experimented with Doc2Query-Llama2~\citep{basnet2024deeperimpact} without notable improvement over Doc2Query-T5.} in Equation~\ref{eq:doc2query}, we use doc2query-T5\footnote{\url{https://huggingface.co/BeIR/query-gen-msmarco-t5-large-v1}}; we generate 100 queries per document ($n=100$ in Equation~\ref{eq:doc2query}); to ensure that the generated ad-hoc query candidates are both diverse and relevant to their source document, we follow~\citep{zhuang2022bridging,nogueira2019doc2query} to adopt a top-$k$ sampling strategy ($k=10$) during the query generation.
We use RankLLaMA~\citep{ma2024fine}\footnote{\url{https://huggingface.co/castorini/rankllama-v1-7b-lora-passage}} as the relevance model (Equations~\ref{eq:filter_qd} and \ref{eq:filter_qc}) for query filtering.
We use summation as the aggregation function in Equation~\ref{eq:filter_overall}.
For each conversational context with multiple relevant documents, we apply the processes in Sections~\ref{sec:method_doc2query} and~\ref{sec:method_qf} to each document, and concatenate the selected queries into a single learning target; further exploration of handling multiple relevant documents is left for future work.
For the \our model (Equations~\ref{eq:loss} and \ref{eq:infer}), we initialise it with \texttt{Mistral-7B-Instruct-v0.3}.
We use the prompts illustrated in Figures~\ref{fig:prompt-context} and \ref{fig:prompt-antici} for the contextualisation and interest anticipation settings, respectively.
We fine-tune the \our model on the training set using QLoRA~\citep{dettmers2023qlora} for one epoch.
All experiments are conducted on 4 NVIDIA A100 GPUs (40GB).
For neural retrievers using queries generated by \our, we set the truncation length of the query encoders to 32 tokens.
%

\begin{table*}[!t]
\caption{
Results of reusing off-the-shelf ad-hoc retrievers under the conversation contextualisation and interest anticipation settings.
Conversational context in the former includes both conversational history and the current user utterance, while in the latter, it consists only of conversational history.
%
``PT'' and ``Inf'' indicate retriever inputs during ad-hoc pre-training and \ac{PSC} inference, respectively; ``Q'' denotes ad-hoc queries; ``Conv'' denotes conversational context; ``Text win'' and ``LMGR'' denote two baselines that convert conversational context into text segments and descriptions, respectively; Conv2Q-1-S/-2-S denote the prompting-only variant of our method; and `` Conv2Q'' denotes our method \our.
The best value in each column is \textbf{bold-faced}.
$^*$ denotes a significant improvement when a retriever uses \our-generated queries at inference, compared to the same retriever with other inputs (paired $t$-test, $p$-value $< 0.05$).
}
\label{tab:zeroshot}
\setlength{\tabcolsep}{1mm}
\resizebox{\textwidth}{!}{%
\begin{tabular}{ll ll cc cc cc cc cc}
\toprule
& &  &  & \multicolumn{6}{c}{ProCIS} & \multicolumn{4}{c}{WebDisc} \\
  \cmidrule(lr){5-10} \cmidrule(lr){11-14} 
& \multirow{3}{*}{Retriever} & \multirow{3}{*}{\makecell[l]{PT}} & \multirow{3}{*}{\makecell[l]{Inf}} & \multicolumn{2}{c}{dev} & \multicolumn{2}{c}{future-dev} & \multicolumn{2}{c}{test} & \multicolumn{2}{c}{val} & \multicolumn{2}{c}{test} \\
 \cmidrule(lr){5-6} \cmidrule(lr){7-8} \cmidrule(lr){9-10} \cmidrule(lr){11-12} \cmidrule(lr){13-14}
&  & & & P@1  & MRR@10  & P@1  & MRR@10  & npDCG@5 & MRR@10 & P@1  & MRR@10   & P@1  & MRR@10 \\
\midrule 
\multirow{12}{*}{\rotatebox{90}{Conversational Contextual.}}&BM25   & - & Conv  & 0.082  & 0.123   & 0.265  & 0.295  & 0.043      & 0.052    & 0.205  & 0.281 & 0.199  & 0.277  \\
&ANCE & Q & Conv  & 0.067 & 0.093 & 0.187  & 0.215  & 0.031 & 0.044 & 0.112  & 0.157 & 0.111 & 0.155  \\
&SPLADE++  & Q & Conv & 0.144  & 0.219 & 0.343 & 0.398 & 0.115 & 0.136   & 0.170   & 0.250 & 0.160 & 0.249 \\
&RepLLaMA & Q & Conv &  0.186  & 0.256    & 0.377   & 0.428 &  0.132    & 0.164  & 0.204 & 0.280  & 0.199 & 0.274 \\
\cmidrule(lr){2-14}
&RepLLaMA  & Q & Text win & 0.187  & 0.252  & 0.401  & 0.452  & 0.139  & 0.174  & 0.225  & 0.297  & 0.218  & 0.291 \\
&RepLLaMA & Q & LMGR & 0.203  & 0.267  & 0.387  & 0.440  & 0.146  & 0.184  & 0.222  & 0.291  & 0.213  & 0.295 \\
\cmidrule(lr){2-14}  
&RepLLaMA & Q & Conv2Q-1-S & 0.311  & 0.385  & 0.459  & 0.522  & 0.261  & 0.368  & 0.234  & 0.302  & 0.232  & 0.314 \\
&RepLLaMA & Q & Conv2Q-2-S & 0.315  & 0.393  & 0.462  & 0.527  & 0.266  & 0.369  & 0.246  & 0.311  & 0.247  & 0.316 \\
\cmidrule(lr){2-14} 
&BM25   & - & Conv2Q & \phantom{1}0.323$^*$  & \phantom{1}0.409$^*$ & \phantom{1}0.399$^*$ & \phantom{1}0.493$^*$ & \phantom{1}0.209$^*$  & \phantom{1}0.288$^*$   & \phantom{1}0.283$^*$ & \phantom{1}0.358$^*$  & \phantom{1}0.274$^*$ & \phantom{1}0.349$^*$ \\
&ANCE   & Q & Conv2Q & \phantom{1}0.434$^*$  & \phantom{1}0.501$^*$ & \phantom{1}0.516$^*$ & \phantom{1}0.576$^*$ & \phantom{1}0.289$^*$    & \phantom{1}0.386$^*$   & \phantom{1}0.284$^*$ & \phantom{1}0.352$^*$  & \phantom{1}0.278$^*$  & \phantom{1}0.347$^*$ \\
&SPLADE++  & Q & Conv2Q & \phantom{1}0.522$^*$  & \phantom{1}0.588$^*$ & \phantom{1}0.612$^*$ & \phantom{1}0.665$^*$ & \phantom{1}0.351$^*$ & \phantom{1}0.477$^*$  & \phantom{1}0.312$^*$ & \phantom{1}0.381$^*$ & \phantom{1}0.302$^*$ & \phantom{1}0.375$^*$ \\
&RepLLaMA   & Q & Conv2Q & \phantom{1}\textbf{0.556}$^*$  & \phantom{1}\textbf{0.611}$^*$ & \phantom{1}\textbf{0.638}$^*$ &  \phantom{1}\textbf{0.685}$^*$ & \phantom{1}\textbf{0.361}$^*$     &  \phantom{1}\textbf{0.494}$^*$   &  \phantom{1}\textbf{0.341}$^*$ & \phantom{1}\textbf{0.417}$^*$  & \phantom{1}\textbf{0.333}$^*$  & \phantom{1}\textbf{0.410}$^*$  \\
\bottomrule

\midrule 
\multirow{12}{*}{\rotatebox{90}{Interest Anticipation}}&BM25   & - & Conv & 0.028   &  0.043  & 0.051   & 0.070  & 0.017     & 0.015  & 0.088  & 0.131 & 0.085  & 0.127  \\
&ANCE & Q & Conv & 0.038 & 0.051 & 0.056 & 0.070  & 0.019 & 0.021  & 0.056  & 0.082  & 0.054  & 0.080   \\
&SPLADE++  & Q & Conv & 0.079 & 0.095  & 0.096 & 0.147 & 0.049 & 0.059  & 0.071  & 0.114  & 0.086 & 0.119  \\
&RepLLaMA  & Q & Conv & 0.090  & 0.126 & 0.137   & 0.183   & 0.053      & 0.064  & 0.098  & 0.143 & 0.096 & 0.141 \\
\cmidrule(lr){2-14} 
&RepLLaMA  & Q & Text win & 0.092  & 0.129  & 0.141  & 0.183  & 0.057  & 0.065  & 0.101  & 0.143  & 0.098  & 0.143 \\
&RepLLaMA & Q & LMGR & 0.098  & 0.133  & 0.145  & 0.191  & 0.059  & 0.070  & 0.105  & 0.151  & 0.102  & 0.149 \\
\cmidrule(lr){2-14} 
&RepLLaMA & Q. & Conv2Q-1-S  & 0.111  & 0.155  & 0.168  & 0.212  & 0.084  & 0.113  & 0.127  & 0.172  & 0.127  & 0.172 \\
&RepLLaMA & Q. & Conv2Q-2-S & 0.113  & 0.159  & 0.171  & 0.216  & 0.087  & 0.116  & 0.130  & 0.175  & 0.130  & 0.175 \\
\cmidrule(lr){2-14} 
&BM25   & - & Conv2Q  &  \phantom{1}0.134$^*$  & \phantom{1}0.174$^*$    &  \phantom{1}0.170$^*$  & \phantom{1}0.214$^*$  &  \phantom{1}0.073$^*$  & \phantom{1}0.102$^*$  & \phantom{1}0.145$^*$ & \phantom{1}0.209$^*$ & \phantom{1}0.145$^*$ & \phantom{1}0.205$^*$ \\
&ANCE   & Q & Conv2Q &  \phantom{1}0.172$^*$   & \phantom{1}0.206$^*$    &  \phantom{1}0.223$^*$   & \phantom{1}0.252$^*$  &  \phantom{1}0.095$^*$   &  \phantom{1}0.121$^*$ & \phantom{1}0.141$^*$ & \phantom{1}0.198$^*$ & \phantom{1}0.142$^*$ & \phantom{1}0.190$^*$ \\
&SPLADE++  & Q & Conv2Q & \phantom{1}0.208$^*$   & \phantom{1}0.238$^*$  & \phantom{1}0.261$^*$ & \phantom{1}0.286$^*$  & \phantom{1}0.101$^*$   & \phantom{1}0.131$^*$  &  \phantom{1}0.151$^*$ & \phantom{1}0.208$^*$  & \phantom{1}0.152$^*$  &  \phantom{1}0.204$^*$   \\
&RepLLaMA & Q & Conv2Q      & \phantom{1}\textbf{0.218}$^*$     & \phantom{1}\textbf{0.248}$^*$    & \phantom{1}\textbf{0.272}$^*$    & \phantom{1}\textbf{0.300}$^*$   &  \phantom{1}\textbf{0.123}$^*$  & \phantom{1}\textbf{0.158}$^*$  & \phantom{1}\textbf{0.166}$^*$ & \phantom{1}\textbf{0.238}$^*$ & \phantom{1}\textbf{0.166}$^*$  & \phantom{1}\textbf{0.235}$^*$ \\
\bottomrule
\end{tabular}%
}
\end{table*}

\begin{table*}[!t]
\caption{
Results of fine-tuning retriever under the conversation contextualisation and interest anticipation settings.
Conversational context in the former includes both conversational history and the current user utterance, while in the latter, it consists only of conversational history.
%
``PT'', ``FT'' and ``Inf'' indicate retriever inputs during ad-hoc pre-training, \ac{PSC} fine-tuning and \ac{PSC} inference, respectively; ``Q'' denotes ad-hoc queries; ``Conv'' denotes conversational context; ``Conv2Q'' denotes ad-hoc queries produced by \our.
The best value in each column is bold-faced.
$^*$ denotes a significant improvement when a retriever uses ad-hoc queries during \ac{PSC} fine-tuning and \ac{PSC} inference, compared to the same retriever using conversational context (paired $t$-test, $p$-value $< 0.05$).
}
\label{tab:finetune}
\setlength{\tabcolsep}{1mm}
\resizebox{\textwidth}{!}{%
\begin{tabular}{ll lll cc cc cc cc cc}
\toprule
& &  &  &  & \multicolumn{6}{c}{ProCIS} & \multicolumn{4}{c}{WebDisc} \\
  \cmidrule(lr){6-11} \cmidrule(lr){12-15} 
& \multirow{3}{*}{Retriever} & \multirow{3}{*}{\makecell[l]{PT}} & \multirow{3}{*}{\makecell[l]{FT}} & \multirow{3}{*}{\makecell[l]{Inf}} & \multicolumn{2}{c}{dev} & \multicolumn{2}{c}{future-dev} & \multicolumn{2}{c}{test} & \multicolumn{2}{c}{val} & \multicolumn{2}{c}{test} \\
 \cmidrule(lr){6-7} \cmidrule(lr){8-9} \cmidrule(lr){10-11} \cmidrule(lr){12-13} \cmidrule(lr){14-15}
&  & & & & P@1  & MRR@10  & P@1  & MRR@10  & npDCG@5 & MRR@10 & P@1  & MRR@10   & P@1  & MRR@10 \\
\midrule 
\multirow{6}{*}{\rotatebox{90}{Conv. Context.}}&ANCE & Q & Conv & Conv & 0.446 & 0.529 & 0.548 & 0.613 & 0.281  & 0.420 &  0.246 & 0.318  & 0.247  & 0.313  \\
&SPLADE++ & Q & Conv & Conv & 0.491 & 0.574 & 0.593 & 0.658 & 0.326  & 0.465 &  0.311 & 0.373  & 0.292  & 0.376  \\
&RepLLaMA & Q & Conv & Conv & 0.520 & 0.603 & 0.623 & 0.687 & 0.355  & 0.494 &  0.340 & 0.402  & 0.321  & 0.405  \\
\cmidrule(lr){2-15}
&ANCE & Q & Conv2Q & Conv2Q & \phantom{1}0.484$^*$  & \phantom{1}0.563$^*$  & \phantom{1}0.579$^*$  &\phantom{1}0.650$^*$  & \phantom{1}0.321$^*$  & \phantom{1}0.457$^*$  & \phantom{1}0.319$^*$  & \phantom{1}0.385$^*$  & \phantom{1}0.308$^*$  & \phantom{1}0.378$^*$  \\
&SPLADE++ & Q &Conv2Q & Conv2Q & \phantom{1}0.552$^*$ & \phantom{1}0.620$^*$ & \phantom{1}0.641$^*$ & \phantom{1}0.697$^*$ & \phantom{1}0.386$^*$ & \phantom{1}0.507$^*$ & \phantom{1}0.352$^*$ & \phantom{1}0.424$^*$ & \phantom{1}0.334$^*$ & \phantom{1}0.411$^*$ \\
&RepLLaMA & Q & Conv2Q  & Conv2Q &  \phantom{1}\textbf{0.588}$^*$  & \phantom{1}\textbf{0.648}$^*$ & \phantom{1}\textbf{0.662}$^*$  & \phantom{1}\textbf{0.717}$^*$ & \phantom{1}\textbf{0.397}$^*$  & \phantom{1}\textbf{0.525}$^*$ & \phantom{1}\textbf{0.383}$^*$ & \phantom{1}\textbf{0.458}$^*$  & \phantom{1}\textbf{0.364}$^*$  & \phantom{1}\textbf{0.445}$^*$ \\
\bottomrule

\midrule 
\multirow{6}{*}{\rotatebox{90}{Inter. Antic.}}&ANCE & Q. & Conv & Conv  & 0.127 & 0.156 & 0.144 & 0.180 & 0.061 & 0.076  & 0.068  & 0.093  & 0.061 & 0.090   \\
&SPLADE++ & Q & Conv & Conv  & 0.154 & 0.211 & 0.189 & 0.241 & 0.080 & 0.101 & 0.105 & 0.174 & 0.100 & 0.171 \\
&RepLLaMA  & Q & Conv & Conv  & 0.188 & 0.244 & 0.225 & 0.277 & 0.106 & 0.131 & 0.160  & 0.229 & 0.162  & 0.229 \\
\cmidrule(lr){2-15} 
&ANCE & Q & Conv2Q & Conv2Q & \phantom{1}0.201$^*$   & \phantom{1}0.234$^*$  & \phantom{1}0.252$^*$  & \phantom{1}0.280$^*$ & \phantom{1}0.123$^*$   & \phantom{1}0.151$^*$   & \phantom{1}0.165$^*$  & \phantom{1}0.229$^*$  & \phantom{1}0.160$^*$  &\phantom{1}0.215$^*$  \\
&SPLADE++  & Q & Conv2Q & Conv2Q & \phantom{1}0.236$^*$  & \phantom{1}0.265$^*$   & \phantom{1}0.291$^*$  & \phantom{1}0.315$^*$  & \phantom{1}0.131$^*$    & \phantom{1}0.160$^*$  &  \phantom{1}0.177$^*$ & \phantom{1}0.238$^*$   & \phantom{1}0.181$^*$   & \phantom{1}0.232$^*$   \\
&RepLLaMA & Q & Conv2Q & Conv2Q & \phantom{1}\textbf{0.244}$^*$   & \phantom{1}\textbf{0.278}$^*$  & \phantom{1}\textbf{0.304}$^*$  & \phantom{1}\textbf{0.334}$^*$ & \phantom{1}\textbf{0.147}$^*$ & \phantom{1}\textbf{0.186}$^*$  & \phantom{1}\textbf{0.197}$^*$  & \phantom{1}\textbf{0.270}$^*$ & \phantom{1}\textbf{0.199}$^*$  & \phantom{1}\textbf{0.270}$^*$  \\
\bottomrule
\end{tabular}
}
\end{table*}

\section{Results}
\label{sec:res}

\subsection{From ad-hoc pre-training to PSC inference}
\label{sec:gap1}

To answer \ref{RQ1}, we present results of off-the-shelf ad-hoc retrievers using conversational context, text windows/descriptions and ad-hoc queries generated by \our and its prompting-only variant (Conv2Q-1-S/-2-S), on ProCIS and WebDisc, under the conversation contextualisation and the interest anticipation settings in Table~\ref{tab:zeroshot}.

We have four main observations.
First, \our-generated queries lead to a significant improvement in retrieval quality for the lexical retriever BM25 compared to using conversational context, across all metrics, evaluation sets and settings.
E.g., \our improves BM25's MRR@10 values by 0.241, 0.198, and 0.236 on the ProCIS dev, future-dev, and test sets under conversation contextualisation, and by 0.131, 0.144, and 0.087 under interest anticipation.
We attribute this improvement to \our's ability to remove noise from raw conversational contexts and generate queries containing keywords that reflect users' implicit information needs.

Second, all ad-hoc neural retrievers using \our-generated queries significantly outperform their counterparts using conversational context across all metrics, evaluation sets and settings. 
For instance, \our improves RepLLaMA's MRR@10 values by 0.355, 0.257 and 0.330 on the ProCIS dev, future-dev, and test sets under conversation contextualisation, and by 0.122, 0.117 and 0.094 under interest anticipation.
The improvement demonstrates that \our effectively adapts off-the-shelf ad-hoc neural retrievers to \ac{PSC} by resolving input mismatches, without requiring retriever fine-tuning on \ac{PSC}.

Third, \our outperforms Text Window and LMGR by a large margin across all settings.
We attribute this to two key advantages of \our:
\begin{enumerate*}[label=(\roman*)]
\item \our generates ad-hoc queries that closely resemble those in ad-hoc search datasets used for ad-hoc retriever pre-training, whereas Text Window and LMGR return text segments or descriptions, causing an input format mismatch.
\item \our captures users’ implicit search intents by learning to generate queries that effectively retrieve the annotated relevant documents for a given conversational context. However, the two baselines lack an effective strategy to ensure their text windows or descriptions accurately match users' implicit information needs.
\end{enumerate*}

Fourth, \our significantly outperforms its prompting-only variant (Conv2Q-1-S/-2-S) in retrieval performance.
We think this is because fine-tuning provides \our with extensive training signals to learn to generate queries that accurately align with the annotated relevant documents for each conversational context, effectively capturing users' implicit information needs.
%

\subsection{From ad-hoc pre-training to PSC fine-tuning}
\label{sec:gap2}

To answer \ref{RQ2}, we examine the performance of further fine-tuning three ad-hoc retrievers (ANCE, SPLADE++ and RepLLaMA) on \ac{PSC} training data using raw conversational context and our generated pseudo ad-hoc queries, on ProCIS and WebDisc, under the conversation contextualisation and interest anticipation settings, in Table~\ref{tab:finetune}.
For further fine-tuning using pseudo ad-hoc queries, we fine-tune each retriever using optimal pseudo ad-hoc queries selected by QF-DC from Doc2Query-generated candidates (see Section~\ref{sec:further_ft}).

We have two main observations. 
First, compared to the results in Tables \ref{tab:zeroshot} in Section \ref{sec:gap1}, we found that ad-hoc neural retrievers using \our-generated queries without fine-tuning on \ac{PSC} achieve comparable or superior retrieval performance to retrievers fine-tuned on \ac{PSC} using conversational context.
This finding reiterates the effectiveness of \our in adapting ad-hoc retrievers to \ac{PSC}, even without retriever fine-tuning.
%

Second, we found that all retrievers fine-tuned on our pseudo ad-hoc query targets and inferred with \our-generated queries significantly outperform those fine-tuned and inferred using raw conversational context across all metrics, datasets, and settings.
%
%
We attribute this improvement to the reduced domain distance between ad-hoc pre-training (source) and \ac{PSC} fine-tuning (target) by consistently using ad-hoc query formats.
As a result, retrievers fine-tuned on our pseudo ad-hoc queries can fully leverage pre-trained ad-hoc knowledge gained during ad-hoc pre-training, leading to more effective transfer learning.

\begin{figure}[ht]
    \centering
    \begin{subfigure}{0.495\columnwidth}
        \includegraphics[width=\linewidth]{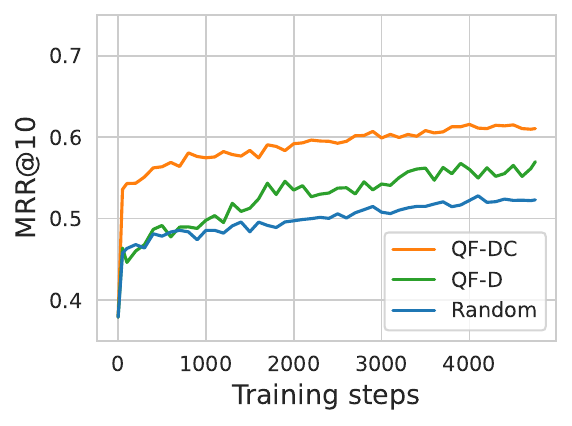}
        \vspace*{-7mm}
        \caption{CC on dev}
        \label{fig:filter_cc_dev}
    \end{subfigure}
    \begin{subfigure}{0.495\columnwidth}
        \includegraphics[width=\linewidth]{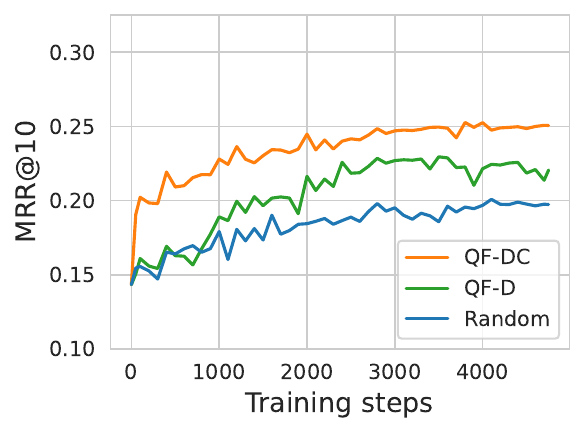}
        \vspace*{-7mm}
        \caption{IA on dev}
        \label{fig:filter_ia_dev}
    \end{subfigure}
    \begin{subfigure}{0.495\columnwidth}
        \includegraphics[width=\linewidth]{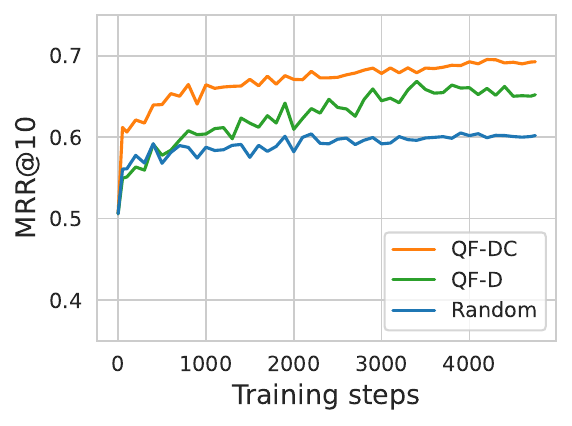}
        \vspace*{-7mm}
        \caption{CC on future-dev}
        \label{fig:filter_cc_future_dev}
    \end{subfigure}
    \begin{subfigure}{0.495\columnwidth}
        \includegraphics[width=\linewidth]{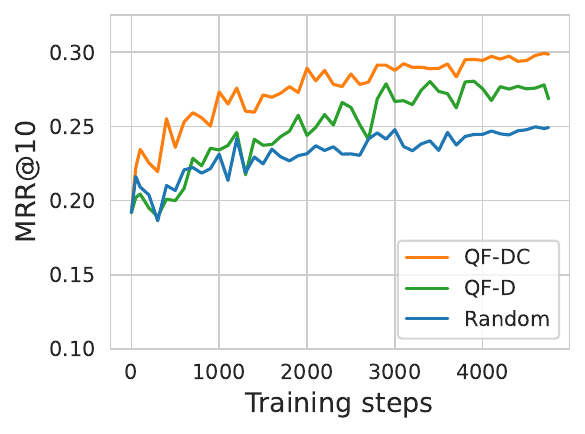}
        \vspace*{-7mm}
        \caption{IA on future-dev}
        \label{fig:filter_ia_future_dev}
    \end{subfigure}
    \begin{subfigure}{0.495\columnwidth}
        \includegraphics[width=\linewidth]{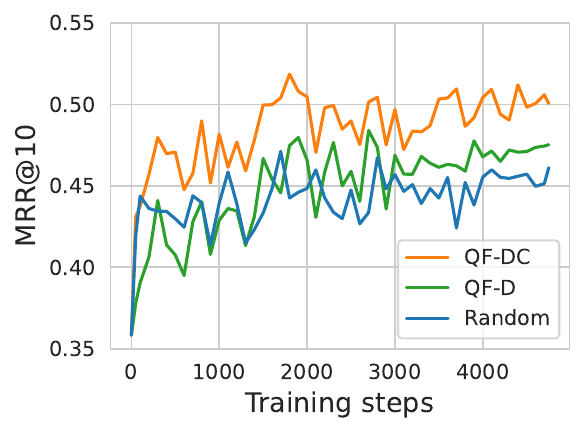}
        \vspace*{-7mm}
        \caption{CC on test}
        \label{fig:filter_cc_test}
    \end{subfigure}
    \begin{subfigure}{0.495\columnwidth}
        \includegraphics[width=\linewidth]{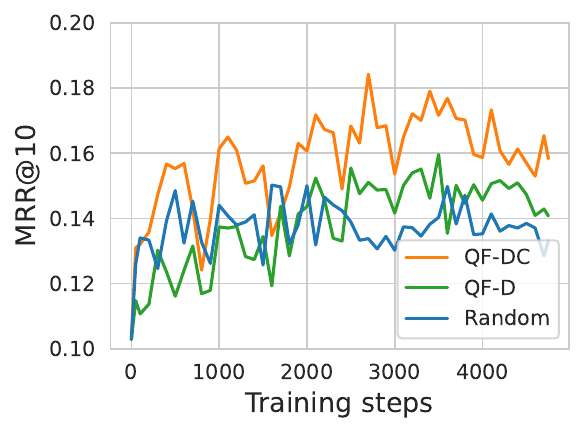}
        \vspace*{-7mm}
        \caption{IA on test}
        \label{fig:filter_ia_test}
    \end{subfigure}
    \caption{
    \our’s learning curves on ProCIS with QF-DC, query–document relevance filtering (QF-D) and no filtering (Random), on the ProCIS dev, future-dev, test sets, under conversational contextualisation (CC) and interest anticipation (IA) settings.
    Each plot shows the retrieval quality (MRR@10) of RepLLaMA using ad-hoc queries generated by \our at different training steps.
    }
    \label{fig:qf}
    \vspace*{1mm}
\end{figure}

\section{Analysis}
\label{sec:ana}

\subsection{Impact of query filtering} 
\label{sec:filter}
To answer \ref{RQ3}, we study the impact of query mechanism by examining \our’s learning curves under three settings: 
\begin{enumerate*}[label=(\roman*)]
\item \textit{QF-DC} (our final approach) uses both query--document relevance and query--conversation alignment;
\item \textit{QF-D} uses query filtering based only on query--document relevance~\citep{gospodinov2023doc2query,jeronymo2023inpars} by removing Equation~\ref{eq:filter_qd} and excluding \smash{$s^{qc}_{t,i}$} in Equation~\ref{eq:filter_overall}; and
\item \textit{Random} randomly selects an ad-hoc query from candidates generated by a Doc2Query model (See implementation details in Section~\ref{sec:setup}).
\end{enumerate*} 
Figure~\ref{fig:qf} presents the retrieval quality (MRR@10) of RepLLaMA (pre-trained on MS MARCO) using \our-generated queries w.r.t. different training steps, with the three query filtering settings; the results are reported on the dev, future-dev, and test sets of ProCIS, under both the conversation contextualisation and interest anticipation settings. 

We have two main observations.
First, QF-D leads to higher retrieval quality than Random.
This suggests that before directly using Doc2Query-generated candidates to fine-tune \our, it is essential to ensure high-quality ad-hoc query learning targets that can effectively retrieve their corresponding source documents.
This finding aligns with previous research on query filtering based on query--document relevance~\citep{gospodinov2023doc2query,jeronymo2023inpars}.
Second, QF-DC results in faster convergence and higher retrieval quality than QF-D.
We think this is because our introduced query--conversation alignment in QF-DC ensures that the selected ad-hoc query learning targets are pertinent to their corresponding conversational context (\our's learning inputs); this reduced semantic gap between inputs and targets enables \our to more effectively learn to generate queries that accurately retrieve their source documents.
Our finding align with previous research~\citep{xing2024rewrite} showing that narrowing the semantic gap between learning inputs and targets facilitates the learning process and leads to better performance.
%

\subsection{Impact of the choice of \acp{LLM}} 
\label{sec:llm}

\begin{figure}[t!]
    \centering
    \begin{subfigure}{\columnwidth}
        \includegraphics[width=\linewidth]{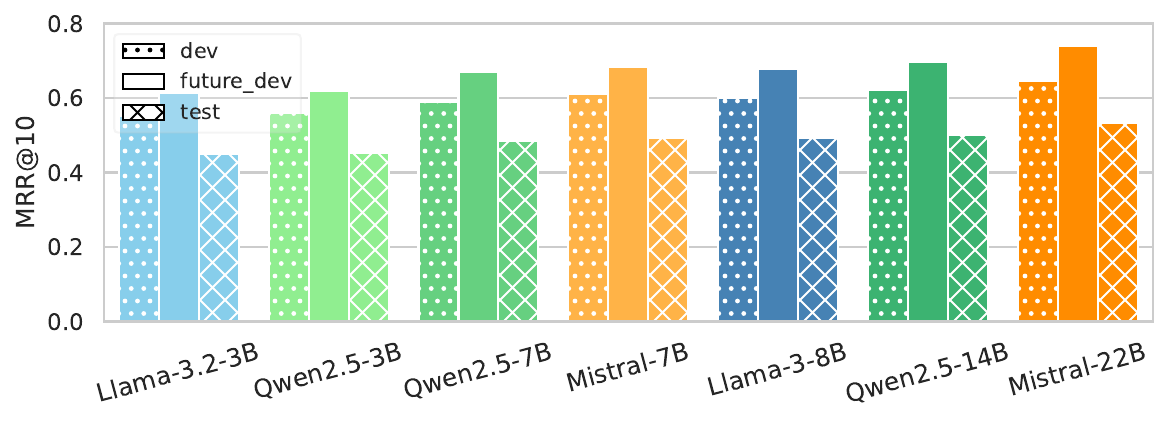}
        \vspace*{-7mm}
        \caption{Conversational contextualisation (CC)}
        \label{fig:llm_cc}
    \end{subfigure}
    \begin{subfigure}{\columnwidth}
        \includegraphics[width=\linewidth]{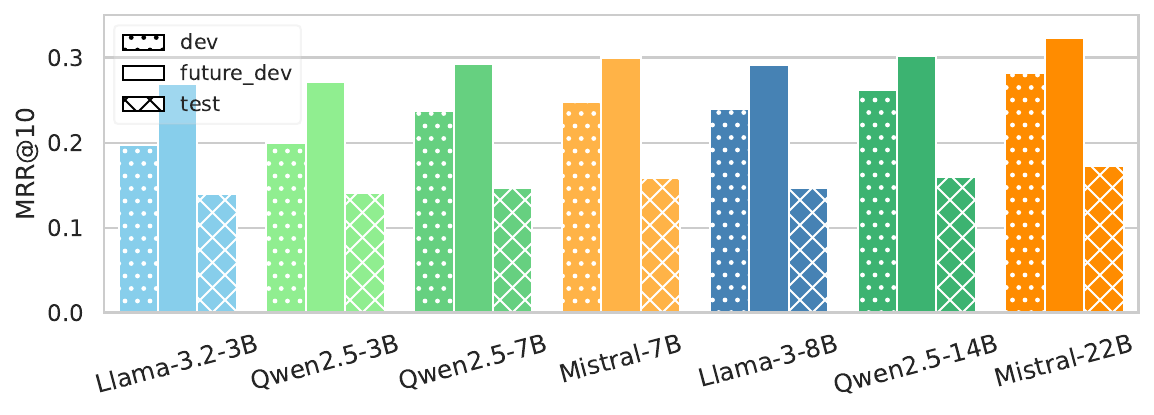}
        \vspace*{-7mm}
        \caption{Interest anticipation (IA)}
        \label{fig:llm_ia}
    \end{subfigure}
    \caption{
    Retrieval quality (MRR@10) of RepLLaMA using ad-hoc queries generated by \our with three families of \acp{LLM} (3B--22B), on the ProCIS dev, future-dev, test sets.
    }
    \label{fig:llm}
    \vspace*{1mm}
\end{figure}

To answer \ref{RQ4}, we examine how the choice of \ac{LLM} impacts \our's performance.
We follow the same fine-tuning setup (see implementation details in Section~\ref{sec:setup}) to evaluate three widely-used \acp{LLM} families, Mistral, Llama and Qwen, spanning from 3B to 22B.
Specifically, for Mistral, we use Mistral-7B-Instruct-v0.3 and Mistral-Small-Instruct-2409~(denoted it as ``Mistral-22B-Instruct''); for llama, we use Llama-3.2-3B-Instruct and Llama-3.1-8B-Instruct; for Qwen, we use Qwen2.5-3B/7B/14B-Instruct.
The results, presented in Figure~\ref{fig:llm}, reveal two key insights.
First, \our performs consistently well across different \ac{LLM} configurations, highlighting its robustness and generalisability across models of varying sizes.
Second, scaling \ac{LLM} size leads to a steady increase in \our's performance.
Mistral-22B, the largest model in our evaluation, results in state-of-the-art retrieval quality.

\section{Conclusions \& Future Work}

We have proposed \our, a novel framework for \ac{PSC}, which effectively adapts ad-hoc retrievers to \ac{PSC} by bridging the input gap between ad-hoc pre-training and \ac{PSC} fine-tuning/inference.
\our learns to map conversational contexts to pseudo ad-hoc query targets that capture users' implicit information needs.
To do so, we have leveraged Doc2Query to generate a set of pseudo queries from relevant documents for each conversational context.
Furthermore, we have devised QF-DC, a novel query filtering mechanism that selects the optimal query target for each conversational context based on document relevance and conversation alignment.
Extensive experimental results have shown that \our significantly improves the performance of ad-hoc retrievers, whether used directly or after fine-tuning.
QF-DC accelerates convergence while improving retrieval performance, and \our remains robust and generalisable across various \ac{LLM} configurations.

Regarding broader implications, \our can be seen as a tool to easily plug any ad-hoc retriever into \ac{PSC}, and can also be applied to scenarios beyond \ac{PSC}, such as query suggestion in \ac{RAG}~\citep{tayal2024dynamic}.

Our work has the following limitations.
First, we focus on what to retrieve in \ac{PSC} without exploring the prediction of retrieval timing. 
Future work might explore using \acf{QPP} methods~\citep{meng2025qpp,meng2025query,arabzadeh2025query,meng2024dc,meng2023query,arabzadeh2024query1,arabzadeh2024query2} for predicting the timing in \ac{PSC}.
%
%
Second, we evaluate our method on two \ac{PSC} datasets based on multi-party Reddit threads, which might not fully represent the diverse scenarios of \ac{PSC}. 
As no better \ac{PSC} datasets are available at the time of writing, it is valuable to curate a more realistic \ac{PSC} dataset in the future.

%
%
%

\subsubsection*{\bf Acknowledgments.}

%
We thank Chen Luo and Jordan Massiah (Amazon) for insightful discussions.
This research was partially funded by the Hybrid Intelligence Center, a 10-year program funded by the Dutch Ministry of Education, Culture and Science through the Netherlands Organisation for Scientific Research, \url{https://hybrid-intelligence-centre.nl}, project nr.\ 024.004.022.

\clearpage

\clearpage
\bibliographystyle{ACM-Reference-Format}
\balance
\bibliography{references}

\end{document}